\documentclass[twocolumn]{aastex61}

\received{July 17, 2018}
\submitjournal{ApJL}

\usepackage{color}                       

\begin{document}

\title{First Detection of Solar Flare Emission in Middle-Ultraviolet Balmer Continuum}

\correspondingauthor{Marie Dominique}
\email{marie.dominique@oma.be}

\author[0000-0002-1196-4046]{Marie Dominique}
\affiliation{Solar-Terrestrial Centre of Excellence -- SIDC, Royal Observatory of Belgium, 3 Avenue Circulaire, 1180 Uccle, Belgium}
\affiliation{Katholiek Universiteit Leuven (KUL), Celestijnenlaan 200b - bus 2400, 3001 Heverlee, Belgium}

\author{Andrei N. Zhukov}
\affiliation{Solar-Terrestrial Centre of Excellence -- SIDC, Royal Observatory of Belgium, 3 Avenue Circulaire, 1180 Uccle, Belgium}
\affiliation{Skobeltsyn Institute of Nuclear Physics, Moscow State University, Leninskie gory, GSP-1, 119991 Moscow, Russia}

\author{Petr Heinzel}
\affiliation{Astronomical Institute, Czech Academy of Sciences, 25165 Ond\v{r}ejov, Czech Republic}

\author{Ingolf E. Dammasch}
\affiliation{Solar-Terrestrial Centre of Excellence -- SIDC, Royal Observatory of Belgium, 3 Avenue Circulaire, 1180 Uccle, Belgium}

\author{Laurence Wauters}
\affiliation{Solar-Terrestrial Centre of Excellence -- SIDC, Royal Observatory of Belgium, 3 Avenue Circulaire, 1180 Uccle, Belgium}

\author{Laurent Dolla}
\affiliation{Solar-Terrestrial Centre of Excellence -- SIDC, Royal Observatory of Belgium, 3 Avenue Circulaire, 1180 Uccle, Belgium}

\author{Sergei Shestov}
\affiliation{Solar-Terrestrial Centre of Excellence -- SIDC, Royal Observatory of Belgium, 3 Avenue Circulaire, 1180 Uccle, Belgium}
\affiliation{Lebedev Physical Institute, Leninskii prospekt, 53, 119991 Moscow, Russia}

\author{Matthieu Kretzschmar}
\affiliation{LPC2E, UMR 7328 Universit\'e d'Orl\'eans and CNRS, 3a av. de la Recherche Scientifique, 45071 Orl\'eans Cedex 2, France}

\author{Janet Machol}
\affiliation{NOAA/NCEI and University of Colorado/CIRES, 325 Broadway, Boulder, CO 80305, USA}

\author{Giovanni Lapenta}
\affiliation{Katholiek Universiteit Leuven (KUL), Celestijnenlaan 200b - bus 2400, 3001 Heverlee, Belgium}

\author{Werner Schmutz}
\affiliation{PMOD/WRC, Dorfstrasse 33, 7260 Davos Dorf, Switzerland}

\begin{abstract}
We present the first detection of solar flare emission at middle-ultraviolet wavelengths around 2000~\AA\- by the channel 2 of the Large-Yield RAdiometer (LYRA) onboard the PROBA2 mission. The flare (SOL20170906) was also observed in the channel 1 of LYRA centered at the H~I Lyman-$\alpha$ line at 1216~\AA, showing a clear non-thermal profile in both channels. The flare radiation in channel 2 is consistent with the hydrogen Balmer continuum emission produced by an optically thin chromospheric slab heated up to 10000 K. Simultaneous observations in channels 1 and 2 allow the separation of the line emission (primarily from the Lyman-$\alpha$ line) from the Balmer continuum emission. Together with the recent detection of the Balmer continuum emission in the near-ultraviolet by IRIS, the LYRA observations strengthen the interpretation of broadband flare emission as the hydrogen recombination continua originating in the chromosphere.

\end{abstract}


\section{Introduction} \label{ch:introduction}

Solar flares and associated coronal mass ejections are the most powerful energy release events in the solar system. Surprisingly little is known about the distribution of the flare energy over the full solar spectrum \citep{2006AdSpR..37.1576V}. Routine measurements of the X-ray and extreme-ultraviolet (EUV) emissions probe only a small part of the total energy radiated during a flare \citep[e.g.][]{2012ApJ...759...71E}. Most of the flare radiation is emitted at longer wavelengths, but observations in this spectral range covering spectral lines and broadband continua are rare \citep{2011A&A...530A..84K, 2016ApJ...816...88K}. The parts of the solar spectrum between 1000 and 3000~\AA, i.e. far ultraviolet (FUV), middle ultraviolet (MUV), and near ultraviolet (NUV), have a probably important but still poorly known contribution to the total energy emitted during flares \citep[e.g.][]{2006JGRA..11110S14W, 2014ApJ...793...70M}. 

Solar spectra at the FUV to NUV wavelengths have been measured by rocket-borne and space-borne experiments \citep{1949ApJ...109....1D, 1968SoPh....3...64B, 2001A&A...375..591C, 2012SoPh..275..115W, 2018A&A...611A...1M}. Semi-empirical quiet-Sun models have been developed \citep{1981ApJS...45..635V, 1993ApJ...406..319F}. As described e.g. by \citet{1971SoPh...18..347G} and \citet{2008uxss.book.....P}, below 1527~\AA\- the quiet Sun spectrum consists of emission continua and emission lines (the strongest line being the H I Lyman-$\alpha$ line at 1216~\AA) and is mostly produced by the chromosphere. Above $\sim$1800~\AA\- the spectrum consists of a number of continua blanketed by numerous absorption lines \citep{1972SoPh...22...64L}, mostly produced by the upper photosphere. The spectrum between 1527~\AA\- and $\sim$1800~\AA\- is an absorption continuum with mostly emission lines, and is produced around the temperature minimum.


The FUV to NUV spectra taken during flares are quite rare \citep{1979ApJ...227..645C, 1984SoPh...90...63L, 1992ApJ...391..393D, 1996ApJ...468..418B}. \citet{2006JGRA..11110S14W} have observed FUV irradiance spectra for four of the largest flares of solar cycle 23; however, with the exception of the Mg~IIk line, the flare signature above 1900~\AA\- was too low to be detected. \citet{2014ApJ...794L..23H} presented the first IRIS \citep[Interface Region Imaging Spectrometer,][]{2014SoPh..289.2733D} measurements of the Balmer continuum during flares in the NUV channel around 2826~\AA. Other, quite rare flare detections in the Balmer continuum were made close to the Balmer recombination edge at 3646~\AA\- by ground-based instruments \citep[e.g.][]{1982SoPh...80..113H, 1983SoPh...85..285N, 2016SoPh..291..779K}. The contributions of the spectral line emission and continua into the total flare radiation may vary strongly, with either line or continuum emission being dominant depending on time and location \citep{2017ApJ...837..160K}. The hydrogen Balmer continuum is produced by the recombination of free electrons generated during strong flare heating in the chromosphere \citep{1986lasf.conf..216A}. The flare emission in the recombination continua is expected to be almost synchronous with the non-thermal hard X-rays bremsstrahlung emission produced by the beam of accelerated electrons \citep[see e.g.][]{2014ApJ...794L..23H}. 

In this Letter, we report the first detection of the solar flare emission in the middle-ultraviolet Balmer continuum, as measured by the Large-Yield RAdiometer (LYRA) onboard the PROBA2 mission.
   
\section{Data description}\label{ch:instrument}
LYRA \citep{2006AdSpR..37..303H, 2013SoPh..286...21D} onboard the PROBA2 mission takes high-cadence (nominally 20 Hz) solar irradiance measurements in four wide spectral channels (see first two columns of Table \ref{tab:flareObs}), two of which are in the FUV and MUV. Channel 1 (also called the Lyman-$\alpha$ channel) takes observations around the Lyman-$\alpha$ line and nearby continua. Channel 2 (historically called "Herzberg channel" due to its relevance to the Herzberg continuum of molecular oxygen in the Earth's atmosphere) observes between 1900 and 2220 \AA. 

\begin{deluxetable*}{lccccc}[b!]
\tablecaption{Characteristics of the X9.3 flare of 2017 September 6 observed by LYRA and GOES. The Lyman-$\alpha$ residual $E'_1$ is obtained from the channel 1 irradiance $E_1$ after subtraction of the contribution of the hydrogen Balmer continuum derived from channel 2 irradiance $E_2$ (see Section \ref{ch:modeling}). $E'_1$ is dominated by the emission in a few strong lines, mostly the Lyman-$\alpha$ and the C lines in the 1200--1550~\AA\- range.}
\tablecolumns{6}
\tablewidth{0pt}
\tablehead{
\colhead{channel} & \colhead{Bandpass,} & \colhead{Pre-flare irradiance,} & \colhead{Peak irradiance (11:58 UT),} & \colhead{Flare increase,}  & \colhead{Flare increase,} \\
\colhead{} & \colhead{\AA\-} & \colhead{erg s$^{-1}$ cm$^{-2}$} & \colhead{erg s$^{-1}$ cm$^{-2}$}  & \colhead{erg s$^{-1}$ cm$^{-2}$} & \colhead{\%}
}
\startdata
\decimals
channel 1 (Lyman-$\alpha$) & 1200 -- 1230\tablenotemark{a} & 6.85 & 6.92 & 0.07 & 0.97 \\
channel 2 (Herzberg) & 1900 -- 2220\tablenotemark{a} & 690.1 & 692.6 & 2.5 & 0.35 \\
channel 3 (Aluminum) & 1 -- 800 & 4.2 & 30.0 & 25.8 & 614 \\
channel 4 (Zirconium) & 1 -- 200 & 1.45 & 25.5 & 24.05 & 1658 \\
Lyman-$\alpha$ residual & 1200--1550 & - & - & 0.05 & - \\
GOES & 1 -- 8 & 0.007 & 1.35 & 1.34 &19185 \\
\enddata
\tablenotetext{a}{The bandpass provided here is as listed in \citet{2013SoPh..286...21D}. See Figure \ref{fig:solar_spectrum} for the detailed spectral transmissions of LYRA channels 1 and 2 that are of importance for this work.}
\end{deluxetable*}\label{tab:flareObs}

\begin{figure*}
\center{\scalebox{1.}{\plotone{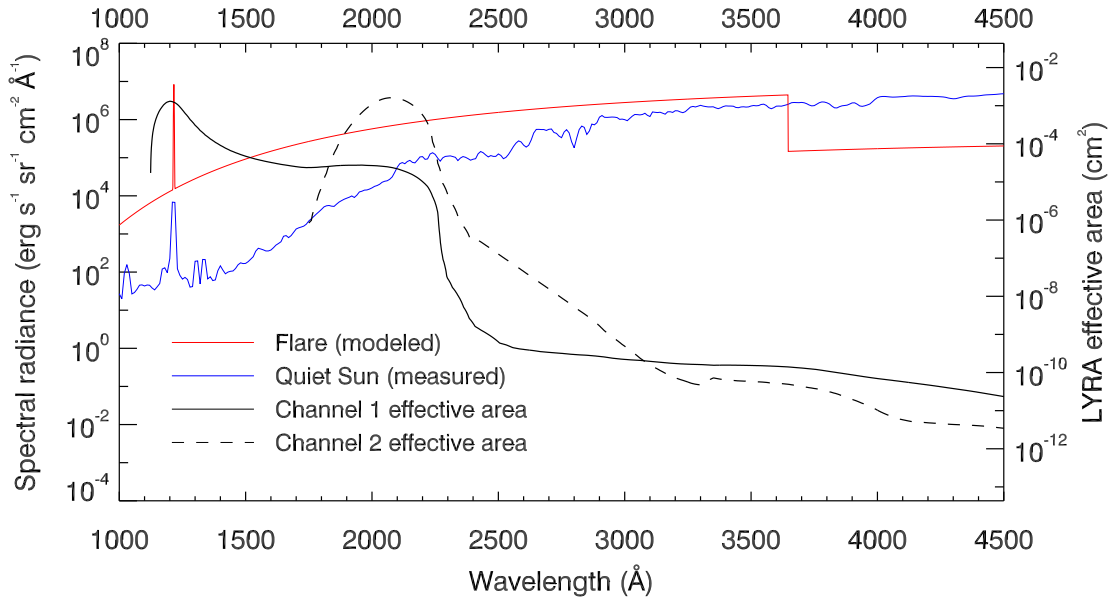}}}
\caption{Solar radiance corresponding to typical quiet-Sun conditions and the increase (without the quiet-Sun background) of radiance produced by the flare. The spectrum $I_\lambda$ (red line) of the flare of 2017 September 6 has been calculated following the procedure described in Section \ref{ch:modeling}. The quiet-Sun spectrum (blue line) measured on 2010 January 7 is shown for comparison. 
The effective areas of the LYRA channels 1 (solid line) and 2 (dashed line) of the spare unit used during the flare campaign are overplotted in black. \label{fig:solar_spectrum}}
\end{figure*}

LYRA was calibrated before the launch at the PTB/BESSY II synchrotron \citep{2013SoPh..286...21D}. The pre-launch effective area for channels 1 and 2 is shown in Figure \ref{fig:solar_spectrum}. One can notice that channel 2 has a high spectral purity, i.e. almost 100\% of the measured signal effectively comes from the 1900 to 2220~\AA\- wavelength range. However, this is not the case for channel 1, for which only 35\% of the measured irradiance comes from the spectral range around Lyman-$\alpha$, while 65\% originates from a plateau in the channel responsivity around 2000~\AA. The latter interval overlaps the spectral range of channel 2, which can be used to disentangle the emission measured in the two channels. The last two channels observe the soft X-rays/EUV range and respectively cover the 1--800 and 1--200~\AA\- intervals.

We also use the data from the 1--8~\AA\- channel of GOES-15 (acquired at a cadence of 2 s), as well as the Solar Dynamics Observatory/Helioseismic and Magnetic Imager (SDO/HMI) continuum images \citep{2012SoPh..275..229S} to determine the surface of the flaring region.
  
\section{Observations}\label{ch:observation}

In September 2017, the NOAA AR 12673 produced 27 M-class flares and 4 X-class flares, among which the two strongest flares observed so far during the solar cycle 24: the X9.3 flare on September 6 and the limb X8.2 flare on September 10.

At the time of these events, LYRA was performing a special flare observation campaign, involving one of its spare unit (i.e. its calibration unit, or unit 1). As this unit was only sporadically opened over the mission, it is relatively well preserved from the ageing process that otherwise affects the instrument \citep{2013SoPh..288..389B}, so it delivered clear observations of the X9.3 flare in all channels. Although about 35\% and 20\% of the sensitivity has been lost since the launch in channels 1 and 2 respectively, the degradation, which is thought to be caused by the deposit of a $\sim$10 nm-thick layer of carbon on the entrance filter, did not modify the spectral characteristics of the instrument.  

The LYRA data set for the X9.3 flare is rather unique. The SXR/EUV channels of LYRA (channels 3 and 4) are specifically used for monitoring solar flares and have captured hundreds of them, but flare observations are relatively rare in channel 1 \citep{2013SoPh..286..221K}. The X9.3 flare was the first flare detected in the channel 2 of LYRA. 

The X8.2 flare, despite being the second strongest flare of the solar cycle, did not produce any signature in LYRA channels 1 and 2. This may be due to the fact that at least one of the footpoints of this flare was located behind the solar limb, hiding the source of the chromospheric emission \citep[see also][]{2018SpaceWeather}. 
Channels 3 and 4, which are the only channels of LYRA measuring coronal emissions, provided clear observations of the flare.     

\begin{figure*}[]
\center{\includegraphics[width=1.\textwidth, trim = {0cm 1cm 0cm 0cm}, clip]{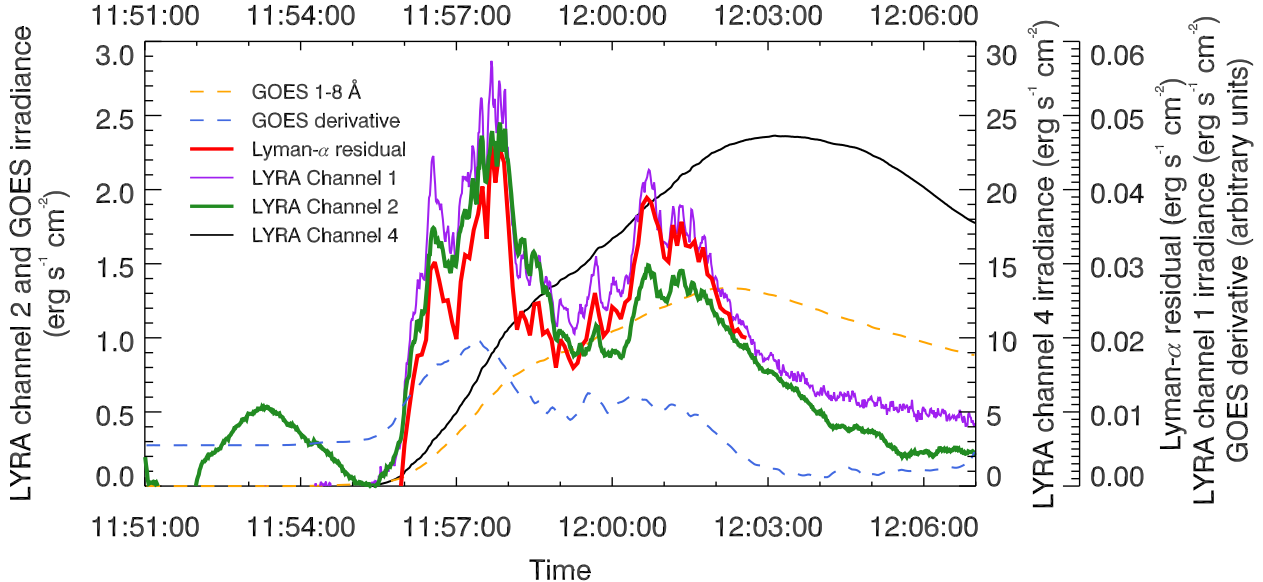}}
\caption{Solar irradiance during the X9.3 flare of 2017 September 6 (with the pre-flare irradiance subtracted), observed by GOES (orange line) and LYRA channels 1, 2 and 4 (respectively the purple, green, and black lines for $E_1$, $E_2$ and $E_4$), as well as the Lyman-$\alpha$ residual irradiance $E'_1$ (red line) extracted from $E_1$. The LYRA data were rebinned to the cadence of 1 s. The time derivative of GOES 1--8~\AA\- data is also shown (blue line) as a proxy of the non-thermal flare emission. Different scales were used for the various time series for the sake of clarity.\label{fig:flare}}
\end{figure*}

The increase of irradiance produced by the X9.3 flare observed by LYRA\footnote{The data in the LYRA channel 3 look very similar to the data taken in channel 4 and are not shown in Figure \ref{fig:flare}.} and by GOES (in the 1-- 8~\AA\- passband) is listed in Table \ref{tab:flareObs} and shown in Figure \ref{fig:flare}. The estimated residual Lyman-$\alpha$ irradiances listed in Table \ref{tab:flareObs} were extracted from LYRA channel 1 following the procedure described in Section \ref{ch:modeling}. In Figure \ref{fig:flare}, the pre-flare irradiance has been subtracted from each timeseries.  

Unfortunately, no hard X-ray measurements are available for this flare. Therefore, we plotted in Figure \ref{fig:flare} the derivative of the GOES data, which constitutes a good proxy for the flare non-thermal emission \citep{1968ApJ...153L..59N}. One immediately sees from Figure \ref{fig:flare} that the emission in channels 1 and 2 looks different from the one in GOES and LYRA channel 4: it is highly modulated and is peaking around 5 minutes earlier. It is similar to the derivative of the GOES 1-8~\AA\- curve. It confirms the non-thermal temporal behavior of emission observed in LYRA channels 1 and 2.   

\section{Spectral Modeling}\label{ch:modeling}

To assess what causes the flare emission in the channel 2 of LYRA, we need to model the flare spectrum around 2000~\AA. Emission in the hydrogen free-bound and free-free continua, in the H$^-$ continuum, as well as in spectral lines, has been considered in the literature \citep{1983SoPh...87..329D, 1985ApJ...299L.103D, 1986lasf.conf..216A,1993ApJ...406..306N, 2014ApJ...783...98K, 2017ApJ...847...48H}. For strong flares in the wavelength range of interest, the emission in the free-bound continuum is expected to be by far the strongest \citep{1986lasf.conf..216A,1993ApJ...406..306N}. 

We therefore adopt the hypothesis that the increase of the irradiance in channel 2 during the flare is primarily due to enhancement of the free-bound continuum of Hydrogen. To calculate the Balmer continuum, we assume that the emission is produced by an optically thin chromospheric slab of plasma with the electron density $n_{\rm e}$ that is enhanced due to increased ionization during the flare. This model was tested e.g. by \citet{1993ApJ...406..306N}, and for Paschen continuum by \citet{2014ApJ...783...98K} and
recently by \citet{2017ApJ...847...48H}. Dominant contribution of the Balmer continuum in MUV and NUV was also predicted by \citet{1986lasf.conf..216A}. The input parameters for a simple slab model are the electron temperature $T$, the electron density $n_{\rm e}$, and the thickness of the emitting layer $L$. The emissivity in the hydrogen recombination continua takes the form \citep{2014tsa..book.....H}:

\begin{equation}
\eta_{\nu}^{i} = n_{\rm e}^2 F_i(\nu, T) \, ,
\end{equation}
where $i$=2 or 3 for Balmer or Paschen continuum, respectively, and $\nu$ is the frequency of the continuum radiation. 
The function $F_i$ is expressed as

\begin{eqnarray}
F_i(\nu, T)  & = & 1.166 \times 10^{14} g(i,\nu) T^{-3/2} B_{\nu}(T) \times \nonumber \\
  &  & e^{h\nu_i/kT} (1 - e^{-h\nu/kT} )/ (i \nu)^3 \, ,
\end{eqnarray}
where $\nu_{i}$ is the continuum-head frequency, 
$g(i,\nu)$ the Gaunt factor, $B_{\nu}(T)$ the Planck function, and $h$ and $k$ are the Planck and Boltzmann constants, respectively \citep{2014tsa..book.....H}. For an optically-thin case, this emissivity is multiplied by $L$ to get the continuum radiance $I_\nu$ (i.e. the specific intensity). Here we assumed an equality between proton and electron densities which is a good approximation in a flaring chromosphere. Furthermore, for the sake of simplicity we took $g(i,\nu)$=1, which is accurate enough for the considered continua. 
Here the continuum radiance $I_\nu$ has units erg s$^{-1}$ cm$^{-2}$ sr$^{-1}$ Hz$^{-1}$, which we convert to $I_\lambda$ in erg s$^{-1}$ cm$^{-2}$ sr$^{-1}$ \AA$^{-1}$ by multiplying $I_\nu$ with the factor 3 $\times$ 10$^{18}$/$\lambda^2$, where $\lambda$ is the continuum wavelength in \AA.

We assumed a typical flare slab temperature of 10000~K. Then, we adjusted the emission measure $n_{\rm e}^2 L$ so that the resulting spectral radiance, once converted into irradiance, multiplied by the instrumental response of the channel 2 of LYRA (see dashed line in Figure \ref{fig:solar_spectrum}) and integrated over its bandpass, provides a result consistent with the channel 2 measurements at any time $t$:
\begin{equation}
E_{2}(t) = C_2  \int_{\lambda} A \frac{I_\lambda(t)}{d^2}S_2(\lambda) \lambda d\lambda
\end{equation}

where $E_{2}$ is the irradiance measured by the channel 2 of LYRA, $d$ is the Sun-Earth distance, $C_2$ is the calibration coefficient of the channel 2, $S_2$ is the effective area of channel 2. $A$ is the emitting area estimated using the method by \citet{2017NewA...57...14M} to be 240~Mm$^2$ at 11:58 UT (the peak time in LYRA channel 2) based on the SDO/HMI observations of the flare in the wing of the Fe I 6173~\AA\- line (M. \v{S}vanda, private communication).


The value of the emission measure $n_{\rm e}^2 L$ producing a spectrum that matches the observations of channel 2 was found to be of $9.1 \times 10^{34} $cm$^{-5}$ at the time of the peak of the flare. \citet{2017ApJ...847...48H} derived the slab thickness around 200 km for a limb flare detected by SDO/HMI. Considering this value as a representative value for the Balmer-continuum formation region, this results in electron densities of the order of $6.7 \times 10^{13} $cm$^{-3}$, consistent with the values found by \citet{1993ApJ...406..306N} and \citet{2014ApJ...783...98K}. Under these conditions, we estimate the optical thickness at 2000~\AA\- to be around 0.015, which confirms the hypothesis of an optically thin slab.

The obtained spectral radiance increase produced by the flare (without the quiet-Sun background) is shown with the red line in Figure \ref{fig:solar_spectrum}. A composite mean quiet-Sun background spectrum is also shown as the blue line for comparison. This spectrum was obtained by merging the full-Sun integrated datasets taken on 2010 January 7 by three spectrometers: TIMED/SEE \citep{2004SPIE.5660...36W} from 1000 to 1300~\AA\-, SORCE/SOLSTICE \citep{1993JGR....9810667R} from 1300 to 3100~\AA\-, and SORCE/SIM (\citet{2005SoPh..230..141H}) from 3100 to 4000~\AA, and converting it into spectral radiance per unit of emitting surface assuming the uniform emission of the quiet Sun disk. 

As was mentioned in Section \ref{ch:instrument}, only 35\% of the irradiance measured by the channel 1 of LYRA at low solar activity comes from the spectral range around Lyman-$\alpha$, while 65\% originates from a plateau in the channel responsivity around 2000~\AA. Once the Balmer continuum spectrum has been calculated based on the measurements of channel 2, its contribution can be subtracted from the channel 1 measurements: 
\begin{eqnarray}
E'_{1}(t) &=& E_{1}(t) - C_1 \int_{\lambda} A \frac{I_\lambda(t)}{d^2} S_{1}(\lambda) \lambda d\lambda. \label{eq:channel1}
\end{eqnarray}

The remaining emission that we call here "Lyman-$\alpha$ residual" consists mostly of the hydrogen Lyman-$\alpha$ emission and a few strong lines, the most prominent of them being the Si III line at 1206~\AA, the C II line at 1335~\AA, the Si IV doublet around 1400~\AA, the Si II line at 1533~\AA, and the C IV doublet at 1548~\AA\- \citep{1986lasf.conf..216A, 2018arXiv180801488S}. According to Table \ref{tab:flareObs}, the Lyman-$\alpha$ residual contributes around 70\% to the total flare emission measured in channel 1 of LYRA.

If the entirety of the remaining signal were attributed to the emission in the Lyman-$\alpha$ line, here modelled by a Gaussian centered at 1216~\AA\- with a 1~\AA\- FWHM (although the line, far from being Gaussian, has extended wings), then the line would be around 500 times more intense than the Balmer continuum, as shown by the peak on the red curve in Figure \ref{fig:solar_spectrum}. It is important to note however that even if the Lyman-$\alpha$ line were responsible for most of the remaining signal, the contribution of other neighbouring lines should not be excluded. In comparison to the line contributions, the emission in the continua around Lyman-$\alpha$ is expected to be small.

\section{Summary and Discussion}\label{ch:discussion}

The X9.3 flare on 2017 September 6 was observed by PROBA2/LYRA in its four channels. This was the first LYRA observation of a solar flare in the MUV wavelengths around 2000~\AA. We demonstrated that the emission detected at these wavelengths by LYRA is consistent with the hydrogen Balmer continuum emission produced by an optically thin chromospheric slab heated up to 10000 K. The densities around $6.7 \times 10^{13}$ cm$^{-3}$ required for the slab thickness of around 200 km are consistent with previous works \citep{1993ApJ...406..306N, 2014ApJ...783...98K, 2017ApJ...847...48H}. Simultaneous observations in channels 1 and 2 of LYRA allow the separation of the line emissions (primarily from the hydrogen Lyman-$\alpha$ line at 1216~\AA) from the Balmer continuum emission generated at longer wavelengths.  

Recently, the Balmer continuum emission from an X1 flare was observed by IRIS around 2826~\AA, as reported by \citet{2014ApJ...794L..23H}. Our radiance at the flare peak computed at 2000~\AA\- is 5.7 $\times$ 10$^5$ erg s$^{-1}$ sr$^{-1}$ cm$^{-2}$ \AA$^{-1}$. Converting this to IRIS NUV we get 2.3 $\times$ 10$^6$ erg s$^{-1}$ sr$^{-1}$ cm$^{-2}$ \AA$^{-1}$, about eight times more as compared to the value given by \citet{2014ApJ...794L..23H} for a weaker X1 flare. We can also convert our radiance to wavelength 6173~\AA\- used by SDO/HMI (i.e. dominated by Paschen continuum), getting the value 2.6 $\times$ 10$^5$ erg s$^{-1}$ sr$^{-1}$ cm$^{-2}$ \AA$^{-1}$. This latter value can be compared with visible-continuum flare detections, but one has to keep in mind that our value of the radiance is the mean value averaged over the flare area. Also a comparison with HMI enhancement may be problematic because during strong flares the HMI 'continuum' signal seems to be strongly contaminated by the flare emission in the Fe I line \citep{2018ApJ...860..144S}.

The contribution of other continua around 2000~\mbox{\AA} \citep[which are usually produced by the quiet photosphere, see ][]{1968SoPh....3...64B} is probably small \citep{1986lasf.conf..216A}. Our value of the peak radiance at 2000~\AA\- is consistent with models of \citet{1986lasf.conf..216A} and lies somewhere between the radiance produced by their F2 and F3 models. This may contribute to better understanding of the physics of white-light flares, although we are detecting enhancements in MUV, not in the white (visible) light. However, the conversion to Paschen-continuum enhancement is a signature of the white-light flare. 

Reports of flares in Lyman-$\alpha$ in the literature \citep[e.g.][]{1984SoPh...90...63L, 2009A&A...507.1005R, 2013SoPh..286..221K, 2017ApJ...848L...8M} are relatively rare and often debated. A recent paper by \citet{2016A&A...587A.123M} questioned the origin of the Lyman-$\alpha$ flare emission reported by broad-band instruments \citep[in particular SDO/EVE,][and LYRA]{2012SoPh..275..115W}, as these detections displayed a thermal-like temporal profile and peaked much later than the non-thermal emission, contrarily to the spectroscopic observation by \citet{1984SoPh...90...63L}. They suggested that these observations might rather correspond to out-of-band emission. LYRA produced very few observations of flares in its Lyman-$\alpha$ channel \citep[channel 1, see][]{2013SoPh..286..221K} due to its fast degradation \citep{2013SoPh..288..389B}. The previous few LYRA observations were all acquired with its nominal or the main backup unit, and they showed a thermal behaviour similar to that described by \citet{2016A&A...587A.123M}. The X9.3 flare on 2017 September 6 is the first flare observed by the channel 1 of the calibration unit, which was better preserved from degradation. 

The temporal correlation of the flare emission measured by LYRA channels 1 and 2 with the GOES derivative confirms that the emission in those channels comes from regions of non-thermal behaviour. The Lyman-$\alpha$ residual irradiance clearly follows a non-thermal profile. It is therefore likely that the anomalous behaviour \citep[reported by][]{2016A&A...587A.123M} of the previous detections by SDO/EVE and in channel 1 of the other two units of LYRA is of instrumental origin (in the case of LYRA, it is probably due to the fast degradation of the nominal unit and the broad spectral range of the main backup unit). 

A limitation of the presented observations is that LYRA integrates the solar flux over the full solar disk and over wavelengths. This does not allow for a clear separation of different continua and spectral lines in the wavelength range of interest (1150--2500~\AA). Spatially and spectrally resolved observations of flares over a wide wavelength range (including the visible light) are necessary to constrain the physics of the broad-band emission in flares \citep{2009AdSpR..43..995V}. 

\acknowledgments

LYRA is a project of the Centre Spatial de Li\`ege, the Physikalisch-Meteorolo\-gisches Observatorium Davos and the Royal Observatory of Belgium funded by the Belgian Federal Science Policy Office (BELSPO) and by the Swiss Bundesamt f\"ur Bildung und Wissenschaft.
M. D., L. D., S. S., and A. N. Z. thank the European Space Agency (ESA) and BELSPO for their support in the framework of the PRODEX Programme. P.H. acknowledges a partial support by the Czech Funding Agency through the grant No. 16-18495S .

\vspace{5mm}
\facilities{PROBA2/LYRA}

\end{document}